\documentclass[conference]{IEEEtran}
\IEEEoverridecommandlockouts
\usepackage{cite}
\usepackage{amsmath,amssymb,amsfonts}
\usepackage{algorithmic}
\usepackage{graphicx}
\usepackage{textcomp}
\usepackage{comment}
\usepackage{hyperref}
\usepackage{flushend}
\usepackage{multirow}
\usepackage{listings}
\usepackage[]{mdframed}
\usepackage[dvipsnames]{xcolor}
\usepackage{fixltx2e}%
\usepackage{diagbox}
\usepackage[linesnumbered,ruled,vlined]{algorithm2e}
\usepackage{amsmath}
\usepackage[most]{tcolorbox}

\def\BibTeX{{\rm B\kern-.05em{\sc i\kern-.025em b}\kern-.08em
    T\kern-.1667em\lower.7ex\hbox{E}\kern-.125emX}}

\newcommand{\rqone}{\textbf{RQ1.} What is the error rate of the automated labeling method based on non-deterministic job reruns? 
}
\newcommand{\rqtwo}{\textbf{RQ2.} How effective is our few-shot learning approach for intermittent job failure detection?}

\newcommand{\rqthree}{\textbf{RQ3.} Can a few-shot model for intermittent job failure detection be used on another project?}

\definecolor{lightgray}{gray}{0.97}

\tcbset{
  summary/.style={
      enhanced,
      sharp corners,
      boxrule=0pt, %
      colback=white,
      interior style={white},
      frame code={
        \draw[black, line width=0.8pt]
          (frame.south west) rectangle (frame.north east);
        \draw[black, line width=0.3pt]
          ([xshift=3pt,yshift=3pt]frame.south west)
          rectangle ([xshift=-3pt,yshift=-3pt]frame.north east);
      },
      left=2pt, right=2pt, top=2pt, bottom=2pt, %
      boxsep=2pt,
  }
}

\begin{document}

\title{Efficient Detection of Intermittent Job Failures Using Few-Shot Learning
}

\author{\IEEEauthorblockN{Henri A\"idasso}
\IEEEauthorblockA{
\textit{École de technologie supérieure}\\
Montreal, Canada \\
henri.aidasso.1@ens.etsmtl.ca}
\and
\IEEEauthorblockN{Francis Bordeleau}
\IEEEauthorblockA{
\textit{École de technologie supérieure}\\
Montreal, Canada \\
francis.bordeleau@etsmtl.ca}
\and
\IEEEauthorblockN{Ali Tizghadam}
\IEEEauthorblockA{
\textit{TELUS}\\
Toronto, Canada \\
ali.tizghadam@telus.com}
}

\maketitle

\begin{abstract}
  One of the main challenges developers face in the use of continuous integration (CI) and deployment pipelines is the occurrence of intermittent job failures, which result from unexpected non-deterministic issues (e.g., flaky tests or infrastructure problems) rather than regular code-related errors such as bugs. Prior studies developed machine learning (ML) models trained on large datasets of job logs to classify job failures as either intermittent or regular. As an alternative to costly manual labeling of large datasets, the state-of-the-art (SOTA) approach leveraged a heuristic based on non-deterministic job reruns. However, this method mislabels intermittent job failures as regular in contexts where rerunning suspicious job failures is not an explicit policy, and therefore limits the SOTA's performance in practice. In fact, our manual analysis of 2,125 job failures from 5 industrial and 1 open-source projects reveals that, on average, 32\% of intermittent job failures are mislabeled as regular.
  To address these limitations, this paper introduces a novel approach to intermittent job failure detection using few-shot learning (FSL). Specifically, we fine-tune a small language model using a few number of manually labeled log examples to generate rich embeddings, which are then used to train an ML classifier. Our FSL-based approach achieves 70-88\% F1-score with only 12 shots in all projects, outperforming the SOTA, which proved ineffective (34-52\% F1-score) in 4 projects. Overall, this study underlines the importance of data quality over quantity and provides a more efficient and practical framework for the detection of intermittent job failures in organizations.
\end{abstract}


\section{Introduction}

Central to modern software development lifecycle, continuous integration and continuous deployment (CI/CD) pipelines (or builds) consist of the automated steps that software source code undergoes to be swiftly delivered to end-users \cite{humble_continuous_2010}. These steps, commonly referred to as pipeline \textit{jobs}, include code compilation, static code analysis, unit and integration testing, container image creation, and deployment into production environments \cite{humble_continuous_2010, aidasso_build_2025}. While successful build executions are ideal, build job failures also play a critical role in ensuring software quality. In principle, they provide developers with early feedback on issues in their submitted code changes (e.g., bugs, test failures, and static analysis errors such as code format issues) \cite{olewicki_towards_2022, hilton_usage_2016}. As a result, when job failures occur, developers assume the presence of regular code-related issues and rely on the job execution logs to promptly diagnose and resolve these issues before they reach production \cite{hilton_usage_2016}.

For TELUS, our industrial partner in the telecommunications field, automated builds ensure rapid and reliable delivery of high-quality software-defined networks (SDNs). Software-Defined Networking is a modern networking approach that abstracts the monitoring and dynamic management of large-scale network traffic into several microservice-based applications (i.e., SDNs) \cite{noauthor_software-defined_nodate}. These software products support critical domains such as phone communications, internet access, connected healthcare, and enterprise networking, which makes their fast and reliable delivery crucial. In this context, intermittent job failures (due to issues typically unrelated to code, such as overloaded servers, unauthorized access, networking issues, or infrastructure problems) pose a significant challenge, undermining the reliability of the CI/CD process and leading to inefficient use of both computational and human resources.

In particular, intermittent job failures often mislead developers, who waste a lot of time and resources identifying and diagnosing them \cite{lampel_when_2021, olewicki_towards_2022}. In fact, 
developers tend to rerun these jobs multiple times in the expectation of a different result, i.e., a transition from \textit{failure} to \textit{success} \cite{lampel_when_2021, olewicki_towards_2022, durieux_empirical_2020, aidasso_diagnosis_2025}. At TELUS, one in five job failures is identified as intermittent upon rerun \cite{aidasso_diagnosis_2025}, and a single job failure is rerun up to 12 times on the same commit before passing. As a result, intermittent job failures and associated job reruns constitute a significant waste of CI resources for the organization. They also involve an inefficient use of developers' time, reducing productivity, and leading to significant delays in software releases \cite{lampel_when_2021, olewicki_towards_2022, moriconi_automated_2022, durieux_empirical_2020}.

The challenge of detecting intermittent job failures is common in industrial settings, where prior studies developed traditional machine learning (ML) models for this purpose \cite{lampel_when_2021, olewicki_towards_2022}. Such models perform a binary classification of job failures to differentiate between \textit{intermittent} and \textit{regular} job failures. In particular, Lampel et al. \cite{lampel_when_2021} leveraged in-house CI telemetry data at Mozilla from over 2 million manually labeled job failures to build an ML classifier. However, telemetry data such as CPU load are not typically available in public or shared CI infrastructure \cite{olewicki_towards_2022}. Moreover, the manual labeling of large datasets is impractical. To tackle these challenges, Olewicki et al. \cite{olewicki_towards_2022} introduced the SOTA approach for intermittent job failure detection at Ubisoft. This approach relies on the similarity in TF-IDF \cite{ramos_using_2003} features of job logs. They also proposed an automated data labeling method based on the heuristic about non-deterministic reruns: {if a job is rerun at least once on the same commit ID and changes outcome, it is labeled as intermittent; otherwise, it is considered regular} \cite{olewicki_towards_2022}.

While non-deterministic job reruns indicate whether a job failure is intermittent, the associated heuristic does not reliably capture all the intermittent job failures.  %
As outlined in the limitations of the SOTA approach \cite{olewicki_towards_2022}, if a developer omits to rerun an intermittent job failure or does not rerun it enough times for it to become successful, the automated labeling method incorrectly labels it as regular.
In fact, this method was used as an alternative to manual labeling performed at Mozilla, based on the key assumption that, at Ubisoft, rerunning suspicious job failures is an explicit adopted policy---thereby minimizing the risk of mislabeling in that context \cite{olewicki_towards_2022}. Consequently, in project contexts where that fundamental assumption does not necessarily hold, such as at TELUS, there is a high likelihood of mislabeling many intermittent job failures as regular ones. If substantial, such mislabeled instances can significantly undermine the performance of the SOTA intermittent failure detection models.

On the other hand, recent improvements in small language models and efficient fine-tuning methods like few-shot learning (FSL) have minimized the need for large labeled datasets and heavy computing resources \cite{tunstall_efficient_2022}. As such, FSL represents an unprecedented opportunity to dispense with the need for potentially erroneous automated labeling of large datasets. Instead, one could manually, thus reliably, label a few job log examples (i.e., \textit{shots}) to build performant models for intermittent job failure detection. To our knowledge, no prior study has (1) evaluated how often the automated labeling method (using the non-deterministic rerun heuristic) mislabels intermittent jobs in different project contexts, or (2) investigated few-shot classification using small language models, as a more efficient alternative to the SOTA for detecting intermittent job failures.

To fill these gaps, we collected data from 16,786 job failures across 5 main TELUS projects and 1 open-source project, and manually labeled 2,125 failed jobs. Using these datasets, we conduct several FSL experiments to answer the following research questions (RQs):

\textbf{\rqone} Our results from manual analysis show that an average of 32\% of job failures (and up to 50\%) are incorrectly labeled as regular failures, when they are actually intermittent. Such high labeling error rates exhibit the limitations of the automated labeling method and cast doubt on the effectiveness of the SOTA approach for detecting intermittent job failures.

\textbf{\rqtwo} Our FSL-based approach achieves F1-scores of 70--88\% in all studied projects, outperforming the state-of-the-art in five of the six projects. As suspected, the SOTA approach proved ineffective in the four projects with the highest mislabeling rates, with significantly lower F1-scores between 34 and 52\%. Its best performance of 89\% is only achieved in the project with minimal labeling error, where our FSL approach still achieves 83\%, suggesting our approach is overall more robust and effective.

\textbf{\rqthree} Cross-project predictions using the FSL models also outperform the SOTA approach in four of the six studied projects, suggesting they can be used in non-critical projects to reduce the number of deployed models. However, in many of the best cases, the cross-project performance remains average (49--70\%) compared to a project-specific FSL model, which is recommended.

The main contributions of this study are \textbf{(1)} an empirical evaluation of the existing automated labeling method's accuracy in different industrial and open-source project contexts, and \textbf{(2)} a novel framework leveraging small language models and few-shot learning for efficient detection of intermittent job failures, achieving F1-scores close to or twice that of the state-of-the-art, using only 12 shots. A replication package \cite{aidasso_artifact_nodate} to foster reuse of our framework is made publicly available.

\section{Background and Related Work}

\subsection{Continuous Integration and Deployment at TELUS}

TELUS is a major Canadian telecommunications and technology company offering a wide array of products and digital services across key sectors such as phone communications, internet access, television, healthcare, agriculture, and more. A significant part of its activities involves developing a wealth of software solutions, including Software-Defined Networks (SDNs) \cite{noauthor_software-defined_nodate}, which are used to dynamically manage wide area network traffic, thereby improving network routing, reliability, and security, while reducing infrastructure costs. At TELUS, SDNs are developed in microservices that communicate with each other and with physical infrastructure and devices.

To achieve fast and reliable software delivery, TELUS relies on a highly distributed and self-hosted CI/CD environment. This mainly includes a self-hosted instance of GitLab \cite{noauthor_gitlaborg_2024}, a unified DevOps platform that integrates both version control system (VCS) and CI Service \cite{noauthor_case_nodate}. Developers are encouraged to frequently trigger the CI/CD pipeline (composed of jobs) by submitting small code changes to GitLab. The pipeline jobs typically include code compilation, static code analysis, unit and integration testing, container image build, image security scanning, and deployment tasks. These jobs, defined in a \textit{build script} configuration file (i.e., \textit{gitlab-ci.yml}), are executed in assigned Kubernetes (k8s) pods created on a self-hosted container platform. This container platform also hosts the shared container registry service, where container images are pulled from or pushed to during job executions. 

In such a distributed environment, pipeline jobs are highly prone to intermittent failures. Indeed, any job execution implies communicating with various services (e.g., GitLab, image scanning service, static code analysis server, container platform, public cloud platforms, etc.). Also, these communications are generally secured with various mechanisms such as virtual private networks, authentication tokens, and SSH keys. As a result, the CI environment has become susceptible to frequent intermittent job failures due to problems related to networking (e.g., broken connections, host resolution failures), infrastructure (e.g., out-of-memory errors, overloaded CPU, pod failures), and authentication (e.g., SSL issues, expired tokens) \cite{aidasso_diagnosis_2025}. Furthermore, the need for testing SDNs' interactions with physical (often unreachable) devices add flaky tests to the root causes of intermittent job failures. While a line of work \cite{lam_idflakies_2019, gyori_nondex_2016, pinto_what_2020, camara_what_2021, alshammari_flakeflagger_2021, pontillo_toward_2021, parry_evaluating_2022} focused on detecting such flaky tests, they only represent a small subset of intermittent job failures at TELUS \cite{aidasso_diagnosis_2025} and in other industrial contexts \cite{lampel_when_2021, olewicki_towards_2022}.

In principle, build job failures are expected to be caused by regular code issues, such as compilation errors, legitimate test failures, or code quality problems \cite{hilton_usage_2016, hassan_tackling_2019}. So, the occurrence of intermittent job failures misleads developers at TELUS and leads to considerable waste, as also reported at Mozilla \cite{lampel_when_2021} and Ubisoft \cite{olewicki_towards_2022}. Because of the inherent non-deterministic nature of intermittent job failures and sometimes very dense and unclear execution logs, developers tend to rerun suspicious job failures in the expectation of a successful outcome \cite{olewicki_towards_2022, aidasso_diagnosis_2025}. While these multiple job reruns (reaching up to 12 reruns of the same job) may ultimately result in success, they constitute a wasteful usage of machine resources and add extra load to already overburdened CI servers \cite{olewicki_towards_2022, aidasso_diagnosis_2025}. In more complex cases, the diagnosis and resolution of intermittent job failures are primarily the responsibility of TELUS' operations and infrastructure teams. However, developers often waste a significant amount of time trying to determine whether a job failure is intermittent, only to find out the problem is beyond their responsibilities, and even their access permissions. 

Therefore, there is a strong need to build intermittent job failure detection models across projects at TELUS. Such models would improve teams' productivity by acting as an initial filter, enabling developers to focus on regular job failures by confidently redirecting intermittent issues to the teams responsible for diagnosing and resolving them. In addition, these models are expected, as in previous studies \cite{olewicki_towards_2022, lampel_when_2021}, to be used as an alternative to costly manual job reruns.

\begin{figure*}[ht]
  \begin{center}
      \includegraphics[width=.7\textwidth]{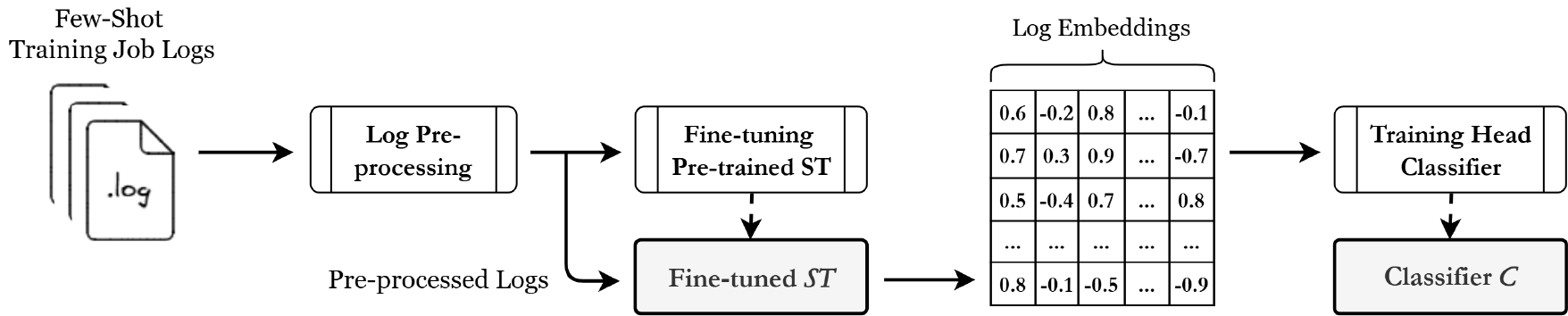}
  \end{center}
\caption{Illustration of the two-step model training following the log pre-processing (\textit{LP}) stage, including, (1) few-shot fine-tuning a sentence transformer \textit{ST} model used to produce embeddings of the training dataset and (2) training a classification head \textit{C} using embeddings generated by the fine-tuned \textit{ST}.}
\label{fig:training}
\end{figure*}

\begin{figure}[!ht]
  \begin{center}
      \includegraphics[width=.49\textwidth]{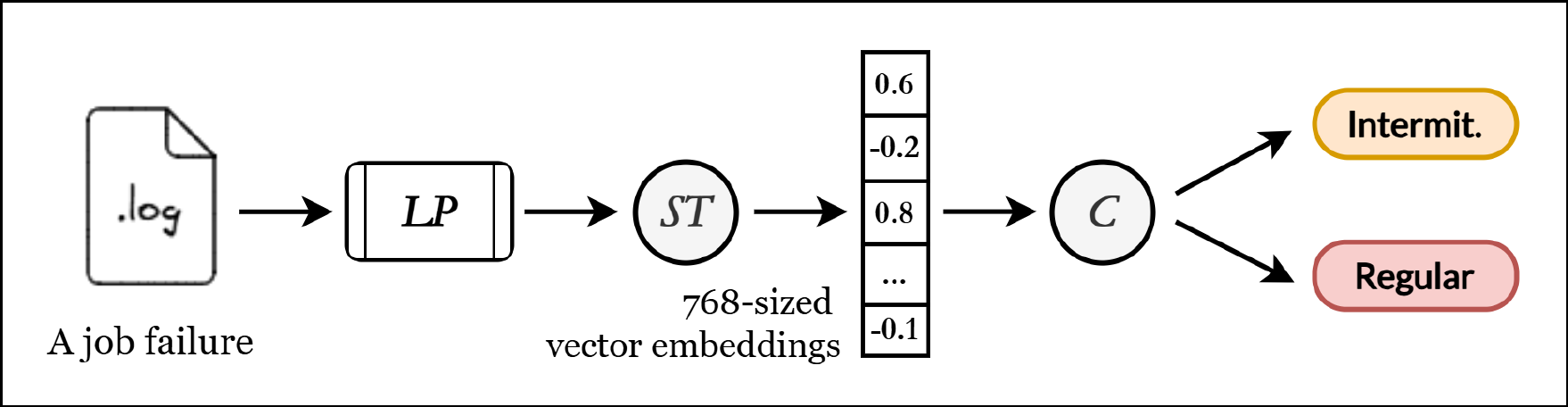}
  \end{center}
\caption{Classifying a job failure as intermittent or not using a few-shot fine-tuned sentence transformer \textit{ST} for embeddings generation and a classifier \textit{C}.}
\label{fig:predicting}
\end{figure}

\subsection{Related Work on Intermittent Job Failures}

Lampel et al. \cite{lampel_when_2021} proposed, to our knowledge, the first approach to language-independent detection of intermittent job failures. This approach relied on job telemetry data, including CPU load and run time as key metrics, from about 2 million jobs at Mozilla. The jobs were manually labeled over time (by a dedicated team of engineers known as \textit{build sheriffs}) and used to train an ML classifier. While this approach achieved good performances at Mozilla, it fails to generalize to other contexts because of two main reasons: (1) the manual labeling of large number of jobs is a costly process to which most organizations including Ubisoft \cite{olewicki_towards_2022}, and also TELUS, are reluctant to dedicate human resources; (2) job telemetry data like CPU load are either not accessible or can not be accurately measured in shared public cloud CI infrastructure due to multitenancy and unknown concurrent changes \cite{olewicki_towards_2022}. 

To address these gaps, Olewicki et al.  \cite{olewicki_towards_2022} introduced the state-of-the-art (SOTA) language-independent approach to detecting intermittent job failures. This ML-based approach leveraged the similarity in TF-IDF \cite{ramos_using_2003} features of job logs and additional metrics, such as the number of prior job reruns. Instead of costly manual labeling, the labeling process was automated using the non-deterministic job rerun heuristic, i.e., a job failure is labeled as intermittent if it was rerun on the same commit and changed its status to success; otherwise, it is considered regular. As a result, the SOTA approach achieved satisfying performances in the context of Ubisoft.

A major limitation of the SOTA approach, as outlined in the study \cite{olewicki_towards_2022}, lies in the fact that its labeling accuracy largely depends on whether developers consistently rerun suspicious job failures, and if so, enough times to exhibit their intermittency. If the rerun practice is not fully and consistently applied, the automated labeling method incorrectly labels the job failure as regular when in fact it is intermittent. One such example from the open-source Veloren project is an intermittent job failure\footnote{\url{https://gitlab.com/veloren/veloren/-/jobs/2720269150}} due to an overloaded server, mislabeled as \textbf{regular} by the heuristic based on non-deterministic reruns. An identical failure\footnote{\url{https://gitlab.com/veloren/veloren/-/jobs/3036439287}} was rerun on another job and passed, enabling the heuristic to assign a correct \textbf{intermittent} label. This highlights the inconsistency of labeling based on the heuristic, which relies too strongly on re-execution practice rather than the root cause of failures. If the number of mislabeled intermittent jobs is high, a model based on log TF-IDF similarity is bound to learn confusing patterns (i.e., similar features in different classes) and therefore perform very poorly. 

Even in contexts where the rerun practice is initially adopted, the automated labeling method limits effective model retraining in the event of concept drift. Indeed, one of the main purposes of using an intermittent job failure detection model is to minimize manual job reruns. So, when such a model needs retraining, there is a high likelihood that recent job failures will not be accurately labelled because of the labeling method's aforementioned limitations. Finally, the existing SOTA approach involves computationally expensive steps for model training, which we discuss further in Section~\ref{sec:validation}.

To address these limitations, we introduce in this paper a novel efficient approach leveraging few-shot learning to create intermittent job failure detection models using only a limited number of manually labeled log examples.

\section{Intermittent Job Failure Detection Using Few-Shot Learning}

Fig.~\ref{fig:predicting} shows an overview of the proposed approach for intermittent job failure detection involving three main stages: (1) log pre-processing (\textit{LP}), (2) embeddings generation using a fine-tuned sentence transformer (\textit{ST})\cite{devlin_bert_2019}, and (3) prediction using the classifier (\textit{C}). Unlike the SOTA \cite{olewicki_towards_2022}, our approach is designed to be more practical, where practitioners only need to provide the job failure logs for the model to classify it as either \textit{intermittent} or \textit{regular}. In addition, the model training process (illustrated in Fig.~\ref{fig:training}) only requires a few examples of job failure logs from each class (i.e., shots). 

To help the model effectively understand log patterns, we first apply a set of log pre-processing rules to reduce noise and minimize log size. For model training, we use SetFit \cite{tunstall_efficient_2022}, an efficient few-shot learning framework with a two-step training process including (1) fine-tuning a pre-trained \textit{ST} model to generate rich text embeddings, and (2) training a classification head model (\textit{C}) using the generated embeddings. At prediction time, the predicted class $c$ of a job failure given its log file is determined as $c = C(ST(LP(log)))$. We present a detailed description of each step in our approach in the following.

\subsection{Log Pre-processing}

Similarly to previous studies \cite{olewicki_towards_2022, jiang_what_2017}, we pre-process the job logs to enable effective learning of the failure patterns. To this end, we leverage the log pre-processing (\textit{LP}) rules described below. These rules, commonly used in natural language processing (NLP), are also inspired by prior studies \cite{olewicki_towards_2022, brandt_logchunks_2020}. We refined them through manual inspections and iterative trials on a random sample of 100 logs. This process ensured that variable elements in the logs were properly abstracted and that noisy elements (e.g., special characters) were cleaned out, while preserving the structure and key information in the logs.

\begin{itemize}
    \item[] \textbf{Rule 1} URLs, file paths, directory paths, durations, and versions identified by regexes are respectively replaced by the constant string \texttt{<URL>}, \texttt{<FILEPATH>}, \texttt{<DIRPATH>}, \texttt{<DURATION>}, and \texttt{<VERSION>}.
    \item[] \textbf{Rule 2} IDs (series of characters containing at least one letter and one number) are replaced by the string \texttt{<ID>}.
    \item[] \textbf{Rule 3} Non letter or numerical characters are removed.
    \item[] \textbf{Rule 4} All numbers, except for HTTP status codes and exit codes, are removed.
    \item[] \textbf{Rule 5} Trailing single letters are removed.
    \item[] \textbf{Rule 6} White spaces and blank lines are removed.
    \item[] \textbf{Rule 7} Duplicate lines are removed.
\end{itemize}

As a result, these rules reduced the size of raw logs by an average of 67\%. We train the models using the pre-processed logs. For prediction, the job logs are also first pre-processed using the same \textit{LP} rules before being passed to the \textit{ST}.

\subsection{Sentence Transformer Few-Shot Fine-tuning}
\label{sec:finetuning}

The model training phase involves two steps. The first step consists of fine-tuning a pre-trained \textit{ST} to produce numerical vector representations called \textit{embeddings} that effectively (i.e., with semantic understanding) distinguish textual logs of the different classes. 
For this purpose, we use {BAAI/bge-small-en-v1.5}\footnote{\url{https://huggingface.co/BAAI/bge-small-en-v1.5}}, the state-of-the-art pre-trained embedding model available on HuggingFace. It maps textual input data to a 768-dimensional dense vector space used for semantic similarity tasks. We chose this language model because it achieves state-of-the-art performance on the MTEB leaderboard \cite{noauthor_baaibge-small-en-v15_2025}, while maintaining a small size (109M parameters), which makes it well-suited for the scalability needs at TELUS. Indeed, small language models enable easy and cost-effective in-house deployments that preserve data confidentiality, unlike larger language models. Moreover, as our results with this small language model were satisfactory, it was unnecessary to explore alternative larger language models in this study.

For \textit{ST} fine-tuning, SetFit leverages a contrastive learning approach \cite{koch_siamese_2015} where the \textit{ST} learns a discriminative embedding space using a dataset of triplets generated from the training set. This dataset includes (1) a set of positive triplets $\{(x_{i}, x_{j}, 1)\}$ where the pairs $x_{i}$ and $x_{j}$ are logs randomly selected from the same class; and (2) a set of the same size composed of negative triplets $\{(x_{i}, x_{j}, 0)\}$ where the pairs $x_{i}$ and $x_{j}$ are logs randomly selected from different classes, i.e., one is \textit{intermittent} and the other \textit{regular}. With this dataset, the pre-trained \textit{ST} model adjusts the embeddings using a cosine similarity loss function so that positive log pairs are positioned closer together in the embedding space. In contrast, negative pairs are pushed far apart. The resulting fine-tuned \textit{ST} model is then used to generate the embeddings of logs in the training set for training the classification head model (see Fig.~\ref{fig:training}). Similarly, the fine-tuned \textit{ST} generates embeddings for each log to be classified at prediction as shown in Fig.~\ref{fig:predicting}.

\subsection{Classification}
\label{sec:classification}

In the second training step, a binary classifier \textit{C} is trained using the training set's embeddings generated by the fine-tuned \textit{ST} model. To this end, we use the \textit{scikit-learn} implementation of the Logistic Regression (LR) algorithm \cite{noauthor_logisticregression_nodate}. This model classifies failed jobs as either \textit{intermittent} (with a predicted value of 1) or \textit{regular} (with a predicted value of 0). 

The validation of our FSL-based approach, covering hyperparameter optimization (HPO) and performance evaluation, is discussed in section \ref{sec:validation}.

\section{Study Design}

\subsection{Research Questions}

The primary goal of this study is to investigate few-shot learning (FSL) for intermittent failure detection, as an innovative and more effective approach that leverages a small set of well-labeled log examples, rather than relying on large datasets that are too expensive to annotate manually and are possibly mislabeled using the existing automated method. Hence, we first assess the labeling error of using the non-deterministic job rerun heuristic for automated labeling (compared to manual labeling) in different project contexts (RQ1). We then evaluate the performance of our FSL-based approach in detecting intermittent job failures, compared to the state-of-the-art approach that relies on large datasets and automated labeling (RQ2). Finally, we investigate the performance of our FSL-based approach in cross-project predictions (RQ3). For this purpose, we design the following research questions (RQs):

\begin{itemize}
    \item[] \rqone
    \item[] \rqtwo
    \item[] \rqthree
\end{itemize}

\subsection{Studied Projects and Data Preparation}
\label{sec:data_preparation}

We conducted this study using build data extracted from five major projects at TELUS and one open-source (OS) project. We focus on these five industrial projects because the study requires tedious manual labeling and consultation with engineers working on the projects.
In addition, we studied the popular open-source role-playing game (RPG) project \texttt{veloren/veloren}\footnote{\url{https://gitlab.com/veloren/veloren}}. We specifically selected this project because the SOTA approach was mainly evaluated on game projects, where intermittent job failures are associated with the heterogeneous nature of the codebase (i.e., containing not only source code but also audio, scene, animation, and data files \cite{olewicki_towards_2022}). In summary, the studied projects are presented in Table~\ref{tab:studied_projects}, with the five industrial projects anonymized for confidentiality reasons. These projects vary in size, maturity, and purpose and encompass 5 programming languages.

\begin{table}[]

\caption{Overview of the Studied Projects and Collected Data}
\label{tab:studied_projects}

\begin{tabular}{|p{.4cm}|l|r|r|r|r|r|r|}
\hline
Proj. & \textbf{\begin{tabular}[c]{@{}l@{}}main lan-\\ guages\end{tabular}} & \multicolumn{1}{l|}{\textbf{\begin{tabular}[c]{@{}l@{}}\#com-\\ mits\end{tabular}}} & \multicolumn{1}{l|}{\textbf{\#yrs}} & \multicolumn{1}{l|}{\textbf{\#jobs}} & \multicolumn{1}{l|}{\textbf{\begin{tabular}[c]{@{}l@{}}\#failed\\ jobs\end{tabular}}} & \multicolumn{1}{l|}{\textbf{BFR}} & \multicolumn{1}{l|}{\textbf{S*}} \\ \hline
A     & TS, Py                                                              & 1.9k                                                                                & 4.9                                   & 7k                                   & 1,597                                                                                 & 21\%                              & 366                              \\ \hline
B     & TS, Py, JS                                                          & 0.7k                                                                                & 3.1                                   & 3k                                   & 1,575                                                                                 & 53\%                              & 346                              \\ \hline
C     & Py                                                                  & 1.7k                                                                                & 3.8                                   & 3k                                   & 1,820                                                                                 & 10\%                              & 347                              \\ \hline
D     & Py, Shell                                                           & 1.6k                                                                                & 1.3                                   & 8k                                   & 2,531                                                                                 & 14\%                              & 369                              \\ \hline
E     & Shell                                                               & 2.9k                                                                                & 1.2                                   & 4k                                   & 2,328                                                                                 & 19\%                              & 331                              \\ \hline
OS    & Rust                                                                & 5.3k                                                                                & 2.5                                   & 57k                                  & 6,935                                                                                 & 22\%                              & 366                              \\ \hline
\end{tabular}
*sample size
\end{table}

We collected job metadata and logs from the projects' build history using the GitLab REST APIs. The collected build jobs data span between 1.2 and 4.9 years of build history across the studied projects. From these jobs, we have only kept completed ones (i.e., with \textit{success} and \textit{failed} statuses) since other non-terminated jobs (e.g., \textit{canceled}) are not relevant for the present study. We obtained a total of 83,214 jobs, including 16,786 failed jobs, split across the projects as indicated in Table~\ref{tab:studied_projects}.

\textbf{Automated Labeling.} We started by labeling intermittent job failures using the non-deterministic job rerun heuristic proposed by Olewicki et al. \cite{olewicki_towards_2022}. %
To this end, we proceeded as follows for each project. First, we grouped the jobs by \textit{name} and \textit{commit} to identify the job rerun groups, i.e., groups containing at least two executions (or runs) of the job on the same commit. Next, we selected rerun groups where the job outcome changed between \textit{failure} and \textit{success}. We assigned the label \textit{intermittent} to the failed jobs within these rerun groups. Olewicki et al. \cite{olewicki_towards_2022} refer to these intermittent job failures as \textbf{brown job failures}, emphasizing their non-deterministic nature of shifting from red failure to green success. We then calculated the Brown Failure Ratio ($BFR = \frac{\#\ brown\ job\ failures}{\#\ total\ job\ failures}$) in each project, and the obtained BFR values are reported in Table~\ref{tab:studied_projects}.

As shown in Table~\ref{tab:studied_projects}, the studied projects present varying BFR values, ranging from 10\% in project \texttt{C} to 53\% in project \texttt{B}. An average \textit{BFR} of 23\% (i.e., in 5 job failures, 1 to 2 are intermittent) across the five TELUS projects, highlights the prevalence of intermittent job failures as a major challenge in these projects. Such a high average BFR also underlines the relevance of focusing on these projects for the present study. Besides, we observe that the BFR of Veloren (22\%) is close to the mean BFR (23\%) of the other studied projects, which suggests its oracle is representative.  We expect the ratios of intermittent job failures to be even higher, as we hypothesize that the non-deterministic rerun method mislabels some of them as regular ones.

\begin{figure*}
  \begin{center}
      \includegraphics[width=.72\textwidth]{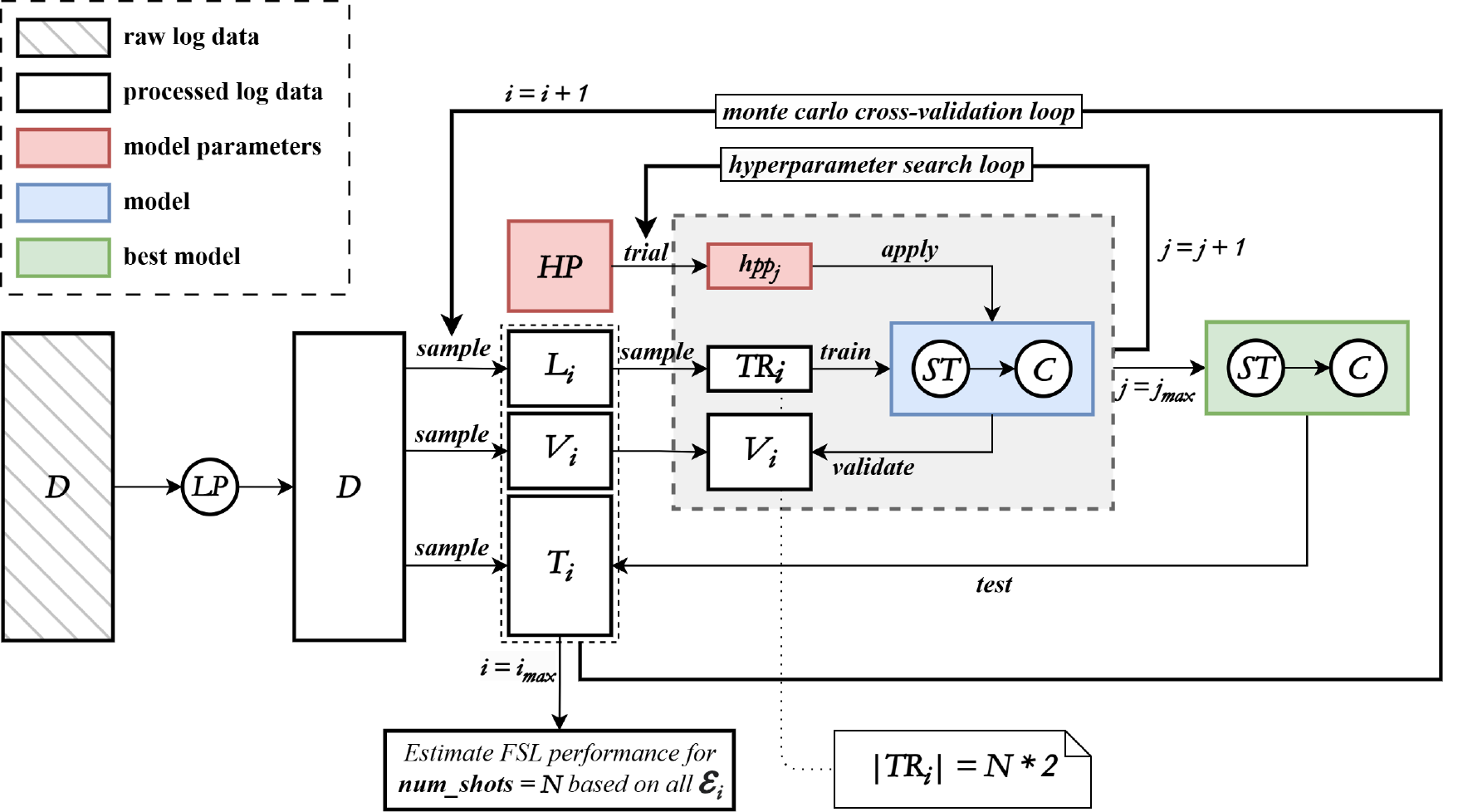}
  \end{center}
\caption{Overview of the experiment for evaluating our FSL-based approach for intermittent job failure detection using a number \textit{N} of shots, i.e., log examples per class. \textit{D} is the manually labeled dataset of logs, and \textit{LP} depicts the log pre-processing step. \textit{L\textsubscript{i}}, \textit{V\textsubscript{i}}, \textit{T\textsubscript{i}}, and \textit{TR\textsubscript{i}}, are the \textit{i\textsuperscript{th}} learning, validation, test, and training set, respectively. \textit{$\varepsilon$\textsubscript{i}} is the model performance estimation on \textit{T\textsubscript{i}}. \textit{HP} represents the hyperparameter space, and \textit{hpp\textsubscript{j}} is the set of hyperparameters for the \textit{j\textsuperscript{th}} loop of hyperparameter search. \textit{ST} represents the sentence transformer that generates log embeddings, while \textit{C} denotes the classifier.}
\label{fig:experiment}
\end{figure*} 

\textbf{Dataset Sampling.} To address the RQs, this study requires manual labeling of failed jobs. Given the impracticality of manually labeling thousands of jobs, we focus on statistically representative samples, following prior work \cite{aidasso_diagnosis_2025, ghaleb_studying_2019}. For each project, we determine the sample size using a confidence level of 95\% and a 5\% margin of error. The obtained sample sizes are indicated in the last column of Table~\ref{tab:studied_projects}.
To maintain the representativeness of the original population, we applied a stratified random sampling so that each sample dataset preserves the same {BFR} as its corresponding original set of failed jobs. Finally, we keep the job logs and labels assigned from automated labeling based on the non-deterministic rerun heuristic. In total, we obtained the execution logs and automated labels of 2,125 failed jobs.

\textbf{Manual Labeling.} The automated labeling method offers the advantage of being inherently reliable when assigning \textit{intermittent} labels. Its potential for mislabeling lies exclusively with job failures that are labeled as \textit{regular} simply because they have not been rerun or have not changed status following reruns. Therefore, our manual labeling process focuses on the job failures initially labeled as \textit{regular}. The manual labeling process of these jobs for each project is described below.

For each job failure initially labeled as \textit{regular} in the sample dataset, we manually analyzed the execution logs to identify the job failure reason using Stack Overflow (SO) \cite{noauthor_stack_nodate}. When the failure reason is suspected to be an intermittent issue, we identify the log chunks that best describe the root cause. We then create regular expressions (regex) to match highly similar logs in the ground truth set of intermittent job logs identified with the automated labeling. In some cases, these regexes enable us to identify similar logs, allowing us to conclude the intermittent nature of the failure initially labeled as regular. 

{However, we noticed that regexes missed several matches due to their rigid nature} (i.e., lacking semantics understanding, and therefore failing to associate \texttt{\small insufficient space} with \texttt{\small out-of-memory} errors, for example). We also observed that many regular failure logs in our industrial projects show symptoms of intermittent issues, without failing because of them, which limits the effectiveness of the regexes. Moreover, we identified a total of 35 regexes during our investigation, which not only fail to reliably match intermittent job failures, but are highly prone to human errors during maintenance.

For these reasons, we ruled out the use of regexes and relied on TELUS engineers' expertise to complete the manual labeling. The failure reasons identified on SO and the associated failure logs have been discussed through weekly meetings with experienced TELUS engineers actively involved in the studied projects and working in different teams (i.e., software development, networking, and infrastructure teams). When a failure reason is finally determined by consensus to be intermittent, the associated job failures (initially automatically labeled as \textit{regular}) are manually relabeled as \textit{intermittent}; otherwise, they are considered to be truly \textit{regular}. 

We saved for each manually labeled job failure, the initial automated label, the manual label, and the justification for relabeling. This includes in some cases, the regex and IDs of matching logs from the ground truth set of intermittent failures, and in other cases, the corresponding intermittent failure category as identified in a recent study \cite{aidasso_diagnosis_2025}. For confidentiality reasons, only data from the OS project are made publicly available in our replication package \cite{aidasso_artifact_nodate}.

\begin{tcolorbox}[enhanced,
      sharp corners,
      boxrule=0pt, %
      colback=white,
      interior style={white},
      frame code={
        \draw[black, line width=0.8pt]
          (frame.south west) rectangle (frame.north east);
        \draw[black, line width=0.3pt]
          ([xshift=3pt,yshift=3pt]frame.south west)
          rectangle ([xshift=-3pt,yshift=-3pt]frame.north east);
      },
      left=.5pt, right=.5pt, top=.5pt, bottom=.5pt, %
      boxsep=2pt,]
\begin{tcolorbox}[summary]
\textbf{Summary of Datasets.} To answer our RQs, we use two datasets per project: the manually labeled samples totaling 2,125 failed jobs with their automated labels, and the remaining dataset totaling 14,661 failed jobs for all projects and labeled only using the automated method.
\end{tcolorbox}
\end{tcolorbox}

\subsection{Validation of the FSL-based approach}
\label{sec:validation}

Fig.~\ref{fig:experiment} illustrates our experimental setup for evaluating our proposed FSL-based approach. The manually labeled dataset \textit{D} is randomly split into learning (\textit{L}), validation (\textit{V}), and test (\textit{T}) sets. The split is stratified so that each set contains the same ratio of intermittent job failures. Similarly to Tunstall et al. \cite{tunstall_efficient_2022}, the learning set is used as the source from which the actual training set (\textit{TR}) is selected using a random sampling. The resulting training set consists of an equal number \textit{N} of shots per class, i.e., \textit{N} regular failure logs and \textit{N} intermittent failure logs. Finally, the validation set \textit{V} is used for HPO, while the test set \textit{T} is used for the model's performance estimation. Algorithm~\ref{alg:fsl_evaluation} shows detailed steps of our evaluation method.

\begin{algorithm}[h]
\label{alg:fsl_evaluation}
\small
\caption{Few-Shot Learning Evaluation (MCCV)}
\KwIn{Dataset $D$, iterations $i_{\max} = 100$, hyperparam. space $HP$, trials $j_{\max}=5$, shots $N$}
\KwOut{Estimated average performance $\bar{\varepsilon}$}

Initialize performance list: $\mathcal{E} \leftarrow []$\;

\For{$i \leftarrow 1$ \KwTo $i_{\max}$}{
    $L_i$, $V_i$, $T_i \leftarrow$  split($D$)\;
    Sample training set, with class(\{$x_k$, $y_k$\}) = \{0, 1\}: $TR_i \leftarrow \left( \{x_k\}_{k=1}^{N} \cup \{y_k\}_{k=1}^{N} \right) \sim L_i$\;
    Initialize: $\text{best\_model} \leftarrow \emptyset$, $\text{best\_score} \leftarrow -\infty$\;

    \For{$j \leftarrow 1$ \KwTo $j_{\max}$}{
        Sample hyperparameters $hpp_j$ from $\mathcal{HP}$\;
        Train model $M_j$ on $TR_i$ using $hpp_j$\;
        Evaluate $M_j$ on $V_i$ and compute score $\varepsilon_{val}^{(j)}$\;
        \If{$\varepsilon_{val}^{(j)} > \text{best\_score}$}{
            $\text{best\_score} \leftarrow \varepsilon_{val}^{(j)}$\;
            $\text{best\_model} \leftarrow M_j$\;
        }
    }

    Evaluate $\text{best\_model}$ on $T_i$ and compute score $\varepsilon_i$\;
    Append $\varepsilon_i$ to $\mathcal{E}$\;
}
Compute average performance: $\bar{\varepsilon} \leftarrow \frac{1}{i_{\max}} \sum_{i=1}^{i_{\max}} \varepsilon_i$\;
\Return{$\bar{\varepsilon}$}
\end{algorithm}

\textbf{Hyperparameter Optimization.} We identified several hyperparameters to tune during the SetFit training steps described in Section~\ref{sec:finetuning} and \ref{sec:classification}. These hyperparameters were selected based on recommendations from the official documentation \cite{noauthor_setfit_nodate}. In particular, we opted for smaller batch size (due to limited memory resources) and lower learning rate ranges following recent guidelines for effective fine-tuning of small language models \cite{pareja_unveiling_2024}. Hence, we define the hyperparameters search space \textit{HP} as follows:

\begin{itemize}
    \item \textit{Body Learning Rate}:
    Learning rate of the \textit{ST} model. \hfill\break
    (Range: $1e^{-6}$ to $1e^{-3}$)
    
    \item \textit{Num Epochs}: Number of training epochs for the \textit{ST}. \hfill\break
    (Choices: 1, 2)

    \item \textit{Batch Size}: Batch size for training the \textit{ST} model. \hfill\break
    (Choices: 2, 4, 8)

    \item \textit{Max Iter}: Maximum number of iterations for the solvers to converge during \textit{C} training. (Range: 50 to 300, by 50)

\end{itemize}

In this study, we run $j_{max} = 5$ trials for hyperparameter search. During each trial, the model (composed of the \textit{ST} and the classifier \textit{C}) is trained using a different set of hyperparameters randomly initialized from the defined \textit{HP} space. After training, each model is evaluated on the validation set $V_{i}$. At the end of the five trials, the model with the highest performance score (i.e., the \textit{best model} out of five) is selected, and its generalization performance is estimated on the test set.

\textbf{Cross-validation Settings.} We use the Monte Carlo Cross-Validation (MCCV) technique \cite{simon_resampling_2007}, also known as repeated random subsampling cross-validation, to estimate the generalization performance of the models for a given number $N$ of shots. This validation technique minimizes the performance estimation error associated with the random selection of the $N * 2$ examples forming the training set. Furthermore, MCCV provides more reliable performance estimates than traditional K-fold CV in binary classification with limited datasets \cite{shan_monte_2022}.

As depicted in Fig.~\ref{fig:experiment}, MCCV consists of repeating model training and performance estimation $i_{max}$ times, each time using different stratified random splits for the learning, validation, and test sets. During our experiments on a 16GB GPU, we found each model training to be fast, from just a few minutes for the smallest number of shots to about 40 minutes for the larger ones. Consequently, we set the number of repeats for MCCV to $i_{max} = 100$ for a given number of shots. In each iteration $i$, the dataset $D$ is split as follows:

\begin{itemize}
    \item Learning set ($L_{i}$): 25\% of the dataset $D$
    \item Validation set ($V_{i}$): 25\% of the dataset $D$
    \item Test set ($T_{i}$): 50\% of the dataset $D$
\end{itemize}

We choose large split sizes for the validation and test sets since (1) we are using limited (manually labeled) sample datasets and (2) our FSL-based approach only requires a few examples for training. Given the sample sizes as indicated in Table~\ref{tab:studied_projects}, the size of learning sets across the projects ranges from 83 to 92, which is large enough for FSL. In fact, the actual training set $TR_{i}$ is created by randomly selecting $N$ intermittent and $N$ regular job failures from the learning set $L_{i}$. Finally, the validation set $V_{i}$ is used to optimize the training parameters of the $ST$ and Logistic Regression $C$ models, while the test set $T_{i}$ is used to estimate the performance $\varepsilon_{i}$ of the best model obtained at the $i^{th}$ iteration.

MCCV is performed once for a given number $N$ of shots and a project. Each cross-validation involves 500 model training (100 repeats and 5 trials per repeat for HPO). As only the best model is retained after HPO, the final performance $\bar{\varepsilon}$ for cross-validation is determined by averaging the 100 $\varepsilon_{i}$ performance values obtained at the end of each iteration.

\textbf{Performance Metrics.} To evaluate the performance of our models, we use the common evaluation metrics, including precision $(\frac{TP}{TP + FP})$, recall $(\frac{TP}{TP + FN})$, and F1-score $(\frac{2}{pre^{-1} + rec^{-1}})$ similarly to prior studies \cite{olewicki_towards_2022, lampel_when_2021}. We report precision and recall scores for each of the \textit{intermittent} and \textit{regular} classes. Nonetheless, the F1-score is used for HPO, and for analysis and comparisons with the SOTA baseline.

\textbf{Baseline.} We use the SOTA approach for intermittent job failure detection proposed by Olewicki et al. \cite{olewicki_towards_2022} as a baseline. For model training, this approach relies on the larger datasets automatically labeled. We removed the manually labeled samples from the datasets, as they are later used for model evaluation. The SOTA model creation involves the following main stages, more extensively described in the original paper. First, we perform log pre-processing and compute the additional required job metrics, including the \texttt{\small number of prior reruns} and the \texttt{\small number of commits since} the last intermittent job failure. In the second stage, we compute the TF-IDF vectors of the processed logs and proceed to feature selection to reduce the vectors' dimensionality---using the SelectKBest algorithm from {sklearn} \cite{noauthor_selectkbest_nodate}. The SOTA detector comprise two eXtreme Gradient Boosting (XGBoost \cite{noauthor_xgboost_nodate}) models: (1) the first model is trained using the best TF-IDF features exclusively, and (2) the second model uses the job metrics and the Shap values \cite{noauthor_welcome_nodate} resulting from the first model training. The final prediction is made through a weighted vote of the two models. In line with the original study, we use a 10-fold cross-validation for hyperparameter optimization. The model with the best hyperparameters is evaluated on the same test sets $T_{i}$ as the corresponding FSL model for comparison. %

\section{Experimental Results}

In this section, we present and discuss the results of our research questions (RQs).

\subsection{\textbf{\rqone}}

\textbf{Motivation.} The goal of this RQ is to evaluate the error rate of the existing automated labeling method based on the non-deterministic job rerun heuristic. Such a finding will help assess the accuracy of the automated labeling method in project contexts without an explicit policy for rerunning suspicious job failures. If found inaccurate, it serves as empirical evidence that further motivates our FSL-based approach for intermittent job detection, since manually labeling large datasets is costly.

\textbf{Approach.} To answer this RQ, we analyze the manually labeled dataset of 2,125 job failure logs composed of representative samples across the six studied projects. These logs are labeled using (1) the automated labeling method based on the non-deterministic rerun heuristic \cite{olewicki_towards_2022} and (2) the manual labeling approach detailed in Section~\ref{sec:data_preparation} and inspired by the study at Mozilla \cite{lampel_when_2021}. The assigned labels are integer values: 1 for \textit{intermittent} and 0 for \textit{regular}. 

Let $a_{i}$ and $m_{i}$ be the assigned automated and manual labels, respectively, to the $i^{th}$ job in a project; and $n$ the total number of jobs in this project. We calculate the automated labeling error rate $E$ in the project as follows:

\begin{equation}
\label{eq:error_rate}
    E = \frac{ \sum_{i=1}^{n} \mid a_{i} - m_{i} \mid }{n}
\end{equation}

The numerator part of Eq~\ref{eq:error_rate} counts mislabeled jobs, i.e., for which $a_{i} \neq m_{i} \implies \mid a_{i} - m_{i} \mid\ =1 $. In addition to the error rate per project, we report statistics about the identified failure reasons (see approach in Section~\ref{sec:data_preparation}) of the intermittent jobs mislabeled as regular by the automated method. Complete artifacts are available in our replication package \cite{aidasso_artifact_nodate}.

\begin{table}[]
\centering
\caption{Statistics and Error rates of Automated labeling compared to Manual labeling across the studied projects.}
\label{tab:labeling_error}
\begin{tabular}{|c|cc|cc|c|}
\hline
\multirow{2}{*}{\textbf{Project}} &
  \multicolumn{2}{c|}{\textbf{Automated}} &
  \multicolumn{2}{c|}{\textbf{Manual}} &
  \multirow{2}{*}{\textbf{\begin{tabular}[c]{@{}c@{}}Error \\ Rate ($E$)\end{tabular}}} \\ \cline{2-5}
              & \multicolumn{1}{c|}{\textbf{Intermit.}} & \textbf{Regular} & \multicolumn{1}{c|}{\textbf{Intermit.}} & \textbf{Regular} &         \\ \hline
A       & \multicolumn{1}{c|}{77}             & 289              & \multicolumn{1}{c|}{175}            & 191              & 26.78\% \\ \hline
B       & \multicolumn{1}{c|}{184}            & 162              & \multicolumn{1}{c|}{240}            & 106              & 16.18\%  \\ \hline
C       & \multicolumn{1}{c|}{36}             & 311              & \multicolumn{1}{c|}{173}            & 174              & 39.48\% \\ \hline
D       & \multicolumn{1}{c|}{52}             & 317              & \multicolumn{1}{c|}{237}            & 132              & 50.14\% \\ \hline
E       & \multicolumn{1}{c|}{64}             & 267              & \multicolumn{1}{c|}{206}            & 125              & 42.90\% \\ \hline
OS      & \multicolumn{1}{c|}{81}             & 285              & \multicolumn{1}{c|}{141}            & 225              & 16.39\% \\ \hline
\end{tabular}
\end{table}

\textbf{Results.} \textbf{On average, one-third (32\%) of job failures---and up to 50\%---are incorrectly labeled as regular by the automated labeling method, when they are actually intermittent. Such high ratios of mislabeled job failures exhibit the limitations of this method and cast doubt on the generalisability of the SOTA detection approach.} Table~\ref{tab:labeling_error} summarizes the automated and manual labeling statistics for each of the six studied projects, along with the calculated error rates. The results show that the mislabeling rates ($E$) vary from one project to another. In the best case, the labeling error rate is at 16.18\% in the project \texttt{B}, meaning that roughly 1 to 2 jobs in 10 is incorrectly labeled in that project. In the worst case, the ratio of mislabeled jobs rises to more than half (50.14\%) of the jobs in the project \texttt{D}. The labeling error rates for the projects \texttt{A}, \texttt{C}, and \texttt{E} are 26.78\%, 39.48\%, and 42.90\%, respectively, resulting in an average error rate of 35.09\% across the five industrial projects. In contrast, the labeling error is at 16.39\% in Veloren, which is nonetheless considerable. Overall, the average labeling error is 31.97\% (i.e., about one-third of the jobs are mislabeled) across the six studied projects.

\begin{table*}[]
\centering
\caption{Performance of FSL models using training sets of $N=$ 12 and 15 examples per class.}
\label{tab:fsl_performances}
\begin{tabular}{|c|lllll|lllll||lllll|}
\hline
\multirow{3}{*}{\textbf{Project}} & \multicolumn{5}{c|}{\textbf{$N$ = 12}}        & \multicolumn{5}{c||}{\textbf{$N$ = 15}}    & \multicolumn{5}{c|}{Baseline SOTA {($N$ = \textit{Full})}}   \\ \cline{2-16} 
 &
  \multicolumn{1}{l|}{\multirow{2}{*}{F1}} &
  \multicolumn{2}{c|}{Intermit.} &
  \multicolumn{2}{c|}{Regular} &
  \multicolumn{1}{l|}{\multirow{2}{*}{F1}} &
  \multicolumn{2}{c|}{Intermit.} &
  \multicolumn{2}{c||}{Regular} &
  \multicolumn{1}{l|}{\multirow{2}{*}{F1}} &
  \multicolumn{2}{c|}{Intermit.} &
  \multicolumn{2}{c|}{Regular} \\
 &
  \multicolumn{1}{l|}{} &
  \multicolumn{1}{l|}{Pre} &
  \multicolumn{1}{l|}{Rec} &
  \multicolumn{1}{l|}{Pre} &
  Rec &
  \multicolumn{1}{l|}{} &
  \multicolumn{1}{l|}{Pre} &
  \multicolumn{1}{l|}{Rec} &
  \multicolumn{1}{l|}{Pre} &
  Rec &
  \multicolumn{1}{l|}{} &
  \multicolumn{1}{l|}{Pre} &
  \multicolumn{1}{l|}{Rec} &
  \multicolumn{1}{l|}{Pre} &
  Rec \\ \hline
A    & \underline{81\textsubscript{5}} & 77\textsubscript{8} & 87\textsubscript{8} & 87\textsubscript{6}  
& 75\textsubscript{12} & \textbf{82\textsubscript{4}} & 78\textsubscript{8} & 88\textsubscript{7} & 88\textsubscript{6} & 76\textsubscript{12}  
& 52 & 37 & 90 & 96 & 62 \\
B    & 83\textsubscript{12} & 90\textsubscript{4} & 79\textsubscript{19} & 69\textsubscript{19} & 78\textsubscript{11} & \underline{84\textsubscript{10}} & 90\textsubscript{4} & 82\textsubscript{17}  & 72\textsubscript{17} & 77\textsubscript{10}  
& \textbf{89} & 86 & 91 & 82 & 72 \\
C    & \underline{70\textsubscript{6}}  & 71\textsubscript{4} & 71\textsubscript{10} & 71\textsubscript{6}  & 71\textsubscript{6} & \textbf{71\textsubscript{6}} & 71\textsubscript{4} & 72\textsubscript{10} & 72\textsubscript{6} & 71\textsubscript{6} & 
34 & 23 & 72 & 91 & 54 \\
D    & \textbf{88\textsubscript{3}}  & 94\textsubscript{3} & 83\textsubscript{6} & 76\textsubscript{6} & 89\textsubscript{6} & \textbf{88\textsubscript{3}} & 94\textsubscript{2} & 83\textsubscript{6} & 75\textsubscript{6} & 90\textsubscript{5} & 
38 & 24 & 97 & 98 & 42 \\
E    & \textbf{80\textsubscript{2}} & 90\textsubscript{5} & 73\textsubscript{5} & 66\textsubscript{3} & 86\textsubscript{8} & \underline{80\textsubscript{3}} & 91\textsubscript{4} & 73\textsubscript{5}  & 66\textsubscript{4} & 87\textsubscript{8}  & 
48 & 33 & 92 & 96 & 47 \\
OS   & \textbf{82\textsubscript{5}} & 91\textsubscript{4} & 75\textsubscript{10} & 86\textsubscript{5} & 95\textsubscript{3} & 
\textbf{82\textsubscript{5}} & 91\textsubscript{3} & 75\textsubscript{9}  & 86\textsubscript{4} & 95\textsubscript{2}  & 
79 & 69 & 95 & 98 & 83 \\ \hline
Median                            
& \textbf{82\textsubscript{5}} & 90\textsubscript{4} & 77\textsubscript{9} & 74\textsubscript{6} & 82\textsubscript{7} 
& \textbf{82\textsubscript{5}} & 91\textsubscript{4} & 79\textsubscript{8}  & 74\textsubscript{6} & 82\textsubscript{7}  &
50 & 35 & 92 & 96 & 58 \\
Mean                              
& \underline{81\textsubscript{6}} & 86\textsubscript{5} & 78\textsubscript{10} & 76\textsubscript{8} & 82\textsubscript{8} & \textbf{81\textsubscript{5}} & 86\textsubscript{4} & 79\textsubscript{9} & 77\textsubscript{7} & 83\textsubscript{7}  & 
57 & 45 & 90 & 94 & 60 \\ \hline

\end{tabular}
\end{table*}

\subsection{\textbf{\rqtwo}}

\textbf{Motivation.} In this second RQ, we investigate whether our few-shot learning approach can effectively detect intermittent job failures with minimal manual labeling, addressing the limitations of the SOTA \cite{olewicki_towards_2022} relying on error-prone automated labeling, as shown in RQ1. In particular, we aim to (1) determine the optimal (i.e., smallest) number of shots required to achieve satisfying performance and to (2) evaluate the relevance in practice of our approach compared to the SOTA for intermittent job failure detection. The findings will guide practitioners in the choice of the optimal number of manually labeled log examples necessary to achieve good and reliable performance, enabling them to apply this FSL-based approach on a larger scale with minimal effort. The results will also offer insights into the effectiveness of our FSL-based approach compared to the SOTA, particularly in projects with no explicit policy for rerunning suspicious job failures.

\textbf{Approach.} We evaluate our FSL-based approach for a given number $N$ of shots using the MCCV technique described in Section~\ref{sec:validation}. We investigate different numbers $N$ of shots ranging from 1 to 15, since above 15 shots, we deviate from the concept of few-shot learning \cite{cao_theoretical_2020}. This maximum of 15 shots (i.e., 30 jobs) for manual labeling has also been discussed with TELUS engineers, who consider it tedious but acceptable. We then conduct the evaluation experiment described in Fig.~\ref{fig:experiment} using the manually labeled sample datasets for each project and number of shots. In total, 45,000 models are trained ($5_{trials} * 100_{repeats} * 15_{shots} * 6_{projects}$), and the resulting performance evaluation measures are saved for analysis. Finally, we analyze the performance evolution of our FSL-based approach as the number of shots is increased, and discuss the choice of the number of shots. Furthermore, we compare in each project the performance of the optimal FSL model with that of the SOTA baseline model also described in Section~\ref{sec:validation}. In line with the original paper \cite{olewicki_towards_2022}, we train and optimize hyperparameters of the SOTA model using the large, automatically labeled datasets. These training sets are deprived of the manually labeled test set splits \textit{T} to avoid set contamination. Hence, the SOTA and our FSL-based approaches are evaluated using the same manually labeled test sets in each project, providing a reliable estimate of their performance in practice.

\begin{figure}
    \advance\leftskip-.15cm
    \centering
    \includegraphics[width=.45\textwidth]{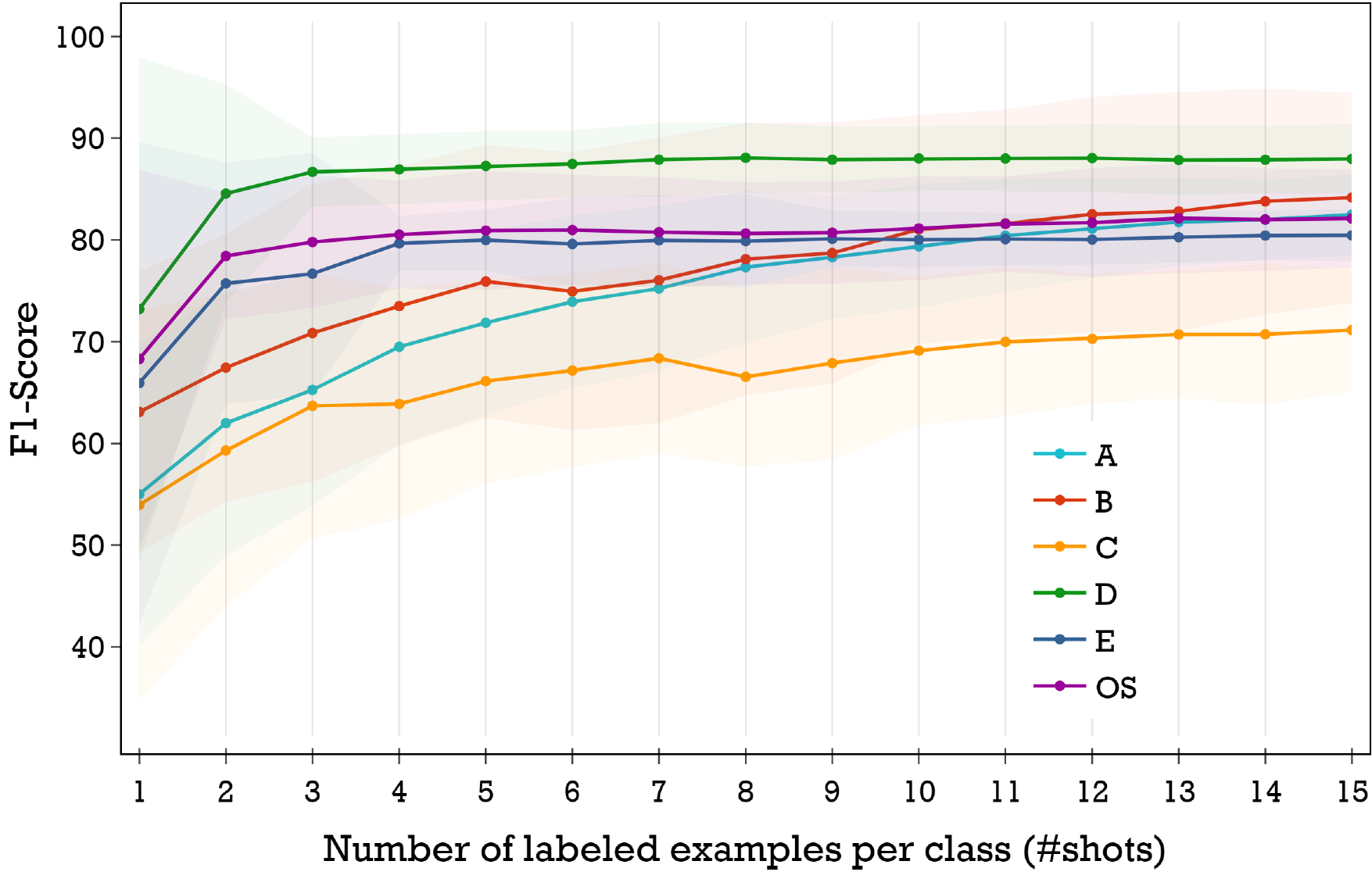}
\caption{Performance of FSL models versus number of manually labeled examples for intermittent job failures detection within studied projects.}
\label{fig:perf_evolution}
\end{figure}

\textbf{Results.} Fig.~\ref{fig:perf_evolution} shows for each project the evolution of the F1-score as the number of shots increases from 1 to 15. It also depicts the standard deviation of F1-scores, using a shaded error band surrounding each line plot. Table~\ref{tab:fsl_performances} presents the detailed performance obtained with 12 and 15 shots, as well as that of the baseline SOTA model trained on full datasets with automated labels. For each project, we highlight in \textbf{bold} the best model performance and \underline{underline} the second best one.

\textbf{Our approach achieves F1-scores above 70\% in all the studied projects (70--88\%), with just 12 shots, demonstrating consistent effectiveness with limited training data. 
Beyond 12 shots, the performance improvement is no longer statistically significant (p-values $>$ 0.05).
} 
As can be observed in Fig.~\ref{fig:perf_evolution}, the model performance tends to improve with a greater number of shots in all projects. However, it exhibits higher variability below 10 shots (i.e., larger error bands from 1 to 9 shots). This is reflected in average F1-scores decreasing in project C between 7 and 8 shots, in project B between 5 and 6 shots, and in the OS project between 6 and 8 shots. Starting from 10 shots, the average performance becomes quite stable and satisfying, with F1-scores at 10 shots ranging between 69\% and 88\% across all the studied projects. With 12 shots, all FSL models achieve average F1-scores exceeding 70\%, specifically 81\%, 83\%, 70\%, 88\%, 80\%, and 82\% in projects A, B, C, D, E, and OS, respectively. Performance further improves and peaks at 15 shots across all projects in our experiments. As reported in Table~\ref{tab:fsl_performances}, the F1-scores at 15 shots remain very close to those at 12 shots, with no to minimal increase of 1\% in projects A, B, and C. 

We investigated whether the small improvement obtained with 15 shots is statistically significant compared with 10 to 14 shots, as we observed a convergence of F1-scores from 10 shots onwards in Fig.~\ref{fig:perf_evolution}. For this purpose, we use the non-parametric Wilcoxon rank-sum test (also referred to as the Mann-Whitney U test \cite{noauthor_mann-whitney_nodate}) to compare the distribution of F1-scores for models trained using 15 shots, with that of models trained using 10, 11, 12, 13, and 14 shots. We obtained respective p-values of 0.003, 0.03, 0.16, 0.45, and 0.78. For 10 and 11 shots, the performance difference is statistically significant with p-values $<$ 0.05 (i.e., 5\% margin of error and confidence level of 95\%). In contrast, starting from 12 shots, all p-values exceed 0.05, indicating no statistically significant difference in performance when training the model with 12 to 14 shots compared to 15. Consequently, we selected 12-shot trained models for all projects, as they achieved the best compromise between performance and number of shots.

\textbf{
FSL models outperform the SOTA models in five of the six projects, with 70--88\% F1-scores versus 34--52\% in four projects. Even in the one exception project, with the lowest labeling error rate, the performance gap is pretty small (83\% vs. 89\%) despite our models only requiring 12 shots}.  As reported in this Table~\ref{tab:fsl_performances}, the SOTA models perform very poorly in projects A, C, D, and E, with respective F1-scores of 52\%, 34\%, 38\%, 48\%. Interestingly, these projects exhibit the highest mislabeling error rates all above 25\% (respectively 26.78\%, 39.48\%, 50.14\%, and 42.90\%) as shown in Table~\ref{tab:labeling_error}. Consequently, the SOTA approach for detecting intermittent failures is not even viable for practical use in these projects. It achieves acceptable performance only in projects B and OS (89\% and 79\% F1-scores, respectively), where we observe the lowest mislabeling error rates of 16.18\% and 16.39\%, respectively. Our FSL-based approach outperforms the SOTA across all the studied projects with just 12 shots, achieving satisfying performance (70--88\% F1-scores).

\subsection{\textbf{\rqthree}}

\textbf{Motivation.} This RQ investigates whether an FSL model trained to detect intermittent failures in one project can be applied to another project. Such \textit{cross-project} models can help practitioners, especially at TELUS with a plethora of active projects, reduce the cost of manual data labeling at scale and avoid deploying a separate model for each project.

\textbf{Approach.} Similarly to RQ2, we evaluate our FSL-based approach using MCCV. For this RQ, we use $N = 12$ shots as the optimal number of shots identified in RQ2. In line with the study at Ubisoft \cite{olewicki_towards_2022}, we follow a one-to-one approach for cross-project predictions, where a model trained on a project is evaluated on each of the other projects.

\textbf{Results.} \textbf{ F1-scores of cross-project predictions using the FSL models range from 18\% to 70\%, outperforming the baseline SOTA models in 4 out of the 6 studied projects. While promising, cross-project performance is still poor to average in many cases and lower than that of project-specific FSL models, which remain recommended.} Table~\ref{tab:cross_project_performances} presents the performance for cross-project predictions. For each project (in line), we report the F1-scores of predictions using models trained on data from the projects in the columns. As \underline{underlined} in this table, cross-project predictions achieve the second-best F1-scores in projects A, C, D, and E (66\%, 62\%, 56\%, and 49\%, respectively), up to two times higher than the SOTA baselines in these projects. Some cross-project combinations are particularly promising. For instance, predictions in project B using project A achieve 70\% F1-score. 

However, cross-project performance is poor in many cases, falling to 18\% when predicting project B using project E. Besides, their predictions are far more variable than those of project-specific models, with standard deviation values reaching up to 22, versus 6 being the highest observed with project-specific models. In conclusion, provided the right job examples, cross-project predictions could be used in some non-critical projects. However, specific models are recommended for better and more reliable performance.

\section{Threats to Validity}

\textbf{Construct Validity.} The main threat to construct validity concerns the correctness of the manual labeling. To mitigate this threat, we define a ground truth set of intermittent job logs determined using the non-deterministic rerun heuristic (i.e., logs of job failures that were rerun at least once on the same commit and transitioned to success). With this ground truth as a reference and input from TELUS engineers, we relabel job failures as intermittent only when their failure reasons are confirmed in consensus after discussions during weekly meetings. Our replication package \cite{aidasso_artifact_nodate} contains (for the OS project) evidence collected during the manual labeling, including similar intermittent jobs in the ground truth set, Stack Overflow resource links and the type of intermittent job failure. These artifacts are being used for further studies on intermittent job failures at TELUS, such as \cite{aidasso_diagnosis_2025}.

\textbf{External Validity.} External validity pertains to the generalizability of our findings. While we conduct this study on industrial projects at TELUS, these projects vary in size, maturity, purpose, programming languages, and intermittent failure ratios. In addition, we studied a popular OS project (Veloren) to support the replication of this study. Unlike the SOTA \cite{olewicki_towards_2022}, the FSL-based approach does not rely on any assumption (i.e., existing policy of rerunning suspicious job failures) and only requires a few (12--15) manually log examples per class to achieve a satisfying performance in all the studied projects. Therefore, our FSL-based approach is expected to generalize to projects with a reasonable log verbosity, regardless of any a priori policy for rerunning job failures. Furthermore, we provide a replication package \cite{aidasso_artifact_nodate} to support the reuse of our approach.

\begin{table}[]
\centering
\caption{Performance of FSL models for Cross-project Predictions.}
\label{tab:cross_project_performances}
\begin{tabular}{|l|rrrrrr|l|}
\hline
\multicolumn{1}{|c|}{\diagbox{Pred}{Train}} & \multicolumn{1}{c}{A}       & \multicolumn{1}{c}{B}       & \multicolumn{1}{c}{C}        & \multicolumn{1}{c}{D}       & \multicolumn{1}{c}{E}       & \multicolumn{1}{c|}{OS}       & $\star$  \\ \hline
A                             & \multicolumn{1}{l}{\textbf{81\textsubscript{5}}} & \multicolumn{1}{l}{\underline{66\textsubscript{9}}} & \multicolumn{1}{l}{47\textsubscript{11}} & \multicolumn{1}{l}{34\textsubscript{6}} & \multicolumn{1}{l}{39\textsubscript{6}} & \multicolumn{1}{l|}{50\textsubscript{11}} & 52 \\ 
B                             & 70\textsubscript{9}                      & \underline{83\textsubscript{12}}                     & 36\textsubscript{18}                      & 48\textsubscript{17}                     & 18\textsubscript{7}                      & 55\textsubscript{3}                       & \textbf{89} \\
C                             & 53\textsubscript{11}                     & \underline{62\textsubscript{9}}                      & \textbf{70\textsubscript{6}}                       & 44\textsubscript{7}                      & 34\textsubscript{3}                      & 49\textsubscript{8}                       & 34 \\
D                             & 46\textsubscript{15}                     & 33\textsubscript{9}                      & 25\textsubscript{13}                      & \textbf{88\textsubscript{3}}                      & 20\textsubscript{2}                      & \underline{56\textsubscript{12}}                      & 38 \\
E                             & \underline{49\textsubscript{17}}                     & 30\textsubscript{6}                      & 23\textsubscript{5}                       & 30\textsubscript{12}                     & \textbf{80\textsubscript{2}}                      & 41\textsubscript{10}                      & 48 \\
OS                            & 42\textsubscript{20}                     & 51\textsubscript{22}                     & 62\textsubscript{16}                      & 48\textsubscript{16}                     & 49\textsubscript{8}                      & \textbf{82\textsubscript{5}}                       & \underline{79} \\ \hline
\multicolumn{8}{l}{$\star$ Baseline SOTA}
\end{tabular}
\end{table}

\section{Conclusion and Future Work}

To tackle the waste associated with intermittent job failures, prior research developed ML techniques to automate their detection. However, these techniques require large datasets of job logs and therefore rely on automated labeling based on too strong assumptions about job reruns. In this study, we found that this automated labeling method mislabels on average 32\% of intermittent job failures as regular. Such high mislabeling rates undermine the performance of the SOTA, with F1-scores capping at 34--52\% in the four projects with the highest (more than 25\%) mislabeling rates. In this paper, we introduced a novel approach for intermittent job failure detection that requires only a few examples of manually labeled job logs per class (i.e., shots) for model training. Our approach achieves 70-88\% F1-scores with just 12 shots, outperforming the SOTA in all projects except one, where it achieves 83\% F1-score versus 89\% for the SOTA. As such, our approach is more effective in practice, while requiring only a limited number of training data. Finally, cross-project predictions appear promising, reaching up to twofold improvement over the SOTA in projects with the highest mislabeling rates. However, these predictions are highly variable and remain average (49--70\%), making project-specific FSL models the recommended choice. In future work, we plan to assess how the choice of training examples influences model performance. As also part of an ongoing work \cite{aidasso_diagnosis_2025}, we intend to explore the detection of the failure categories (i.e., multi-class classification) to support the teams in automated diagnosis and repair of these failures.

\section*{Acknowledgment}

We acknowledge the support of the Natural Sciences and Engineering Research Council of Canada (NSERC), ALLRP/ 576653-2022. This work was also supported by TELUS and Mitacs through the Mitacs Accelerate program.

\bibliographystyle{IEEEtran}
\bibliography{references}

\end{document}